\pdfoutput=1
\documentclass{ws-procs975x65}
\usepackage{microtype}
\usepackage{graphicx}
\usepackage[
unicode=true,
naturalnames=true,
pdfauthor={Xiao-Yong Jin},
pdftitle={Lattice QCD with 12 Quark Flavors: A Careful Scrutiny},
pdfsubject={A discussion about QCD with 12 quark flavors computed with lattice gauge theory},
pdfkeywords={Lattice QCD; Many-flavor-QCD; Walking theory; Conformal theory}
]{hyperref}
\usepackage{bookmark}

\newcommand{\figsubcap}[1]{\par\noindent\centering{\footnotesize #1}}

\begin{document}

\title{Lattice QCD with 12 Quark Flavors: A Careful Scrutiny}

\author{Xiao-Yong Jin}

\address{RIKEN Advanced Institute for Computational Science,\\
  Kobe, Hyogo 650-0047, Japan\\
  E-mail: \href{mailto:xjin@riken.jp}{xjin@riken.jp}}

\author{Robert D. Mawhinney}

\address{Department of Physics, Columbia University,\\
  New York, NY 10027, USA\\
  E-mail:
  \href{mailto:rdm@physics.columbia.edu}{rdm@physics.columbia.edu}}

\begin{abstract}
  With a substantial amount of simulations, we have explored the
  system across a wide range of lattice scales.  We have located a
  lattice artifact, first order bulk transition, have studied its
  properties, and found that the flavor-singlet scalar meson mass
  vanishes at the critical endpoint.  We will discuss the lattice
  phase diagrams and the continuum limits for both a S$\chi$SB phase
  and an IR conformal phase, and compare results with other groups.
\end{abstract}

\keywords{Lattice QCD; Many-flavor-QCD; Walking theory; Conformal
  theory}

\bodymatter

\section{Introduction}

It is both phenomenologically and theoretically interesting to study
strongly coupled gauge theories with the number of massless fermions,
$N_f$, tuned close to the lower limit of the conformal window, where
the theory remains asymptotically free at high energies while
developing a non-trivial infrared (IR) fixed point.  While the system
with $N_f$ close to the upper limit of the conformal window can be
studied perturbatively\cite{Banks:1981nn}, for systems with $N_f$
close to the lower limit, the behavior is less clear due to the
non-perturbative nature of gauge theories.

We have been studying the SU(3) gauge theory with 8 and 12 fundamental
fermions on the lattice, using the naive staggered fermion action and
the DBW2 gauge action.  We have explored the system through basic
low-energy QCD observables with multiple volumes.  In this paper, we
will concentrate on our recent findings of the lattice artifact bulk
transition with 12 flavors, discuss the continuum limit, and compare
results with other groups.  We quote results from largest volume
available at each set of parameters, with finite volume effects of
about a few percent.

\section{Lattice critical point}

We have located a line of first order bulk
transitions\cite{Jin:2012dw} in the bare lattice parameter space:
quark mass, $m_q$, and coupling, $\beta=6/g^2$.  This line of first
order transitions appears at relatively strong lattice couplings
$\beta \sim 0.46$ and small input quark masses $m_q \lesssim 0.008$.
Discontinuities in hadronic observables at the first order transition
become smaller, as the quark mass, $m_q$, is increased.  The system
shows a quick cross-over region\cite{Jin:2009mc}, at $m_q \ge 0.01$,
with a marked change in lattice scales, from strong couplings to
weaker couplings.

\begin{figure}
  \centering
  \includegraphics{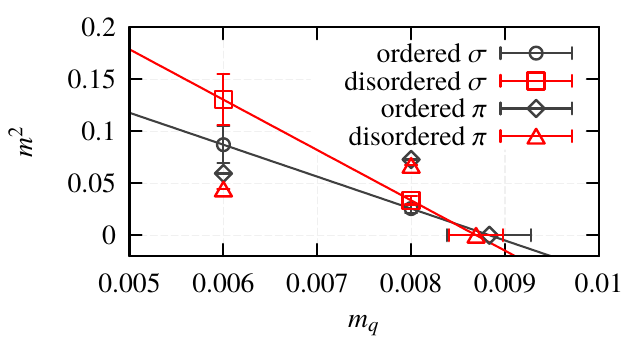}
  \caption{\label{critical_end_point}The critical endpoint from
    extrapolating the mass of the flavor-singlet scalar meson}
\end{figure}

The system is approaching a second order transition, or a critical
endpoint, as the quark mass, $m_q$, is increased while the lattice
coupling, $\beta$, is tuned toward the end of the first order bulk
transition.  We have measured the mass of the flavor-singlet scalar
meson, $\sigma$, using metastable ensembles at two sets of input
parameters, detailed in Ref.~\refcite{Jin:2012dw}.  The mass of the
flavor-singlet scalar meson, $m_\sigma$, depicted in
\fref{critical_end_point}, decreases on the line of the first order
bulk transition, as $m_q$ increases, and becomes even smaller than
pion.  To fit the critical behavior, we have used the mean-field
scaling law, $m_{\sigma} \propto (m_q - m_q^C)^{1/2}$, which happens
to describe our data quite well.  The fitted location of the endpoint
from ordered-start configurations (weak coupling) agrees with the
fitted result from disordered-start configurations (strong coupling).

This critical behavior of $m_\sigma$, resembles the behavior found in
finite temperature transitions\cite{Liao:2001en}.  However, our
findings with 12 quark flavors at zero temperature occur at finite
$m_q$ and $\beta$.  While the continuum physics dictated by asymptotic
freedom, can only be extracted at the limit of $m_q \rightarrow 0$ and
$\beta \rightarrow \infty$, this critical point certainly does not
have such a conventional continuum limit, and is irrelevant to the
continuum physics we are after.  Nonetheless, this critical point does
possess a continuum limit, since the $\sigma$ correlation length
diverges at this point, while all other hadronic scales decouple from
the theory.  This second order endpoint thus gives a scalar field
theory, and is likely non-interacting.\footnote{We don't have hard
  proof, but it is hard to construct a nontrivial scalar field
  theory.}

The first order bulk transition and critical endpoint\cite{Jin:2012dw}
almost surely depends on the lattice action, as a few other groups
have found bulk
behaviors\cite{Cheng:2011ic,Deuzeman:2011pa,Fodor:2012et} with
different lattice actions and bare parameters.  In the parameter space
we have simulated, however, we did not observe the same
characteristics of the bulk transition those groups advertised.  We
acknowledge that certain kind of lattice artifacts are highly likely
to exist in different lattice actions under different disguise and
with different degrees of severity.  Our findings of the critical
behavior of the flavor-singlet scalar meson suggest that caution must
be exercised in interpreting results from the scalar channel on the
lattice, especially when asymptotic scaling is not well established.

\section{Continuum limit}

The continuum limit of lattice gauge theories can be analyzed by the
renormalization group equation of any physical observable,
$P(a,g,\hat{m})$, which depends on lattice spacing, $a$, bare lattice
coupling,\footnote{Not to be confused with renormalization beta
  functions, we use $\beta$ and $g$ interchangeably as the lattice
  bare coupling for ease of discussion.} $g$, and bare lattice input
quark mass, $\hat{m}=ma$.  Near the continuum limit, $P$ should mildly
depend on $a$,
\begin{equation}
  a\frac{\mathrm{d}}{\mathrm{d}a} P =
  \left\{
    a\frac{\partial}{\partial a} -
    \beta(g, \hat{m})\frac{\partial}{\partial g} -
    \gamma(g, \hat{m})\frac{\partial}{\partial \hat{m}}
  \right\} P(a, g, \hat{m}) =
  \mathcal{O}(a)\,, \label{eq:RG}
\end{equation}
where $\beta(g,\hat{m})$ and $\gamma(g,\hat{m})$ are lattice version
of the renormalization group beta functions, and the right hand side
of the equation is the usual scaling error.\footnote{The size of the
  scaling violation depends on the lattice action and observable.}  In
an asymptotically free theory, to find the lattice beta functions, one
would perturbatively compute some observable $P$, and plug the
perturbative form back to \eref{eq:RG}.  See
Ref.~\refcite{Kogut:1982ds} for a massless derivation using the quark
potential.  One will find the beta functions coincide with the
continuum ones up to the lowest universal orders.\footnote{Minus signs
  in \eref{eq:RG} ensure this agreement.}  In an asymptotically free
theory, this procedure is self-consistent, because $g\rightarrow0$,
together with a fine-tuned $\hat{m}\rightarrow0$, leads to
$a\rightarrow0$ according to the beta function calculated from this
procedure.

Away from the physical continuum limit, one can supposedly use a form
of the observable $P$ that contains lattice higher order operators,
say $P \rightarrow P + P_l$, where $P_l$ is irrelevant in the usual
continuum limit of physical QCD.  With this additional term $P_l$,
solving \eref{eq:RG} has the possibility to lead to lattice beta
functions, $\beta_l(g, \hat{m})$ and $\gamma_l(g, \hat{m})$, that are
completely different from the usual continuum QCD, and contain lattice
artifacts.  One can solve these new lattice beta functions for $a
\rightarrow 0$, and obtain a new continuum limit at $g \rightarrow
g_C$ and $\hat{m} \rightarrow \hat{m}^C$, which is different from the
usual continuum limit of physical QCD at $g \rightarrow 0$ and
$\hat{m} \rightarrow 0$.  As long as the result of observable $P+P_l$
is consistent with calculations in the limit of $g \rightarrow g_C$
and $\hat{m} \rightarrow \hat{m}^C$, such that \eref{eq:RG} is
self-consistent, we arrive at a lattice artifact continuum limit.

\begin{figure}
  \centering
  \parbox{0.485\textwidth}{\includegraphics{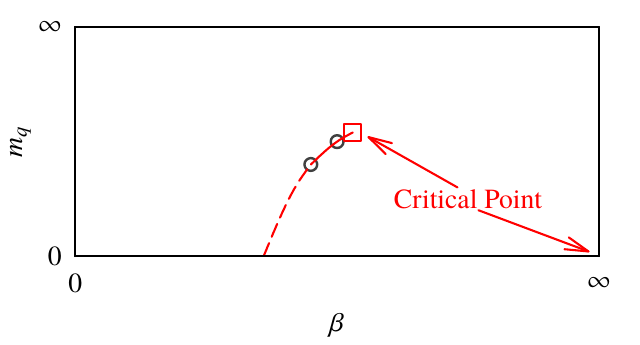}
    \figsubcap{(a) Empirical diagram with 12 flavors}}
  \hfill
  \parbox{0.485\textwidth}{\includegraphics{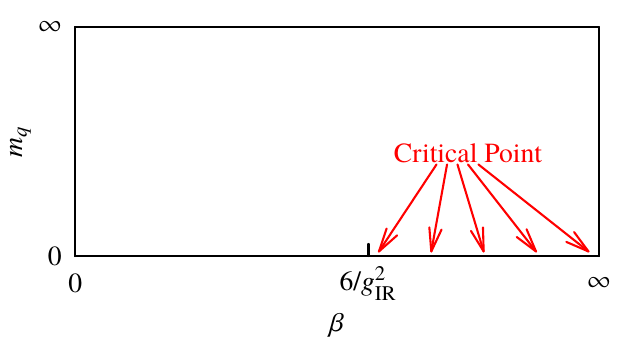}
    \figsubcap{(b) Infinite volume diagram with IR conformal}}
  \caption{\label{bulk_trans}Lattice phase diagrams}
\end{figure}

The second order critical endpoint we found with 12 flavors at finite
lattice coupling $\beta$ and input quark mass $m_q$, presented in the
previous section, is such a lattice artifact continuum limit.  The
empirical lattice phase diagram is sketched in \fref{bulk_trans}(a),
where the axes are lattice bare parameters.  We have two data points,
which are located on the transition line and extended to smaller
masses with a dashed line.  We note two critical points that
correspond to the lattice artifact continuum limit and the physical
QCD continuum limit, respectively.

To verify whether the 12-flavor theory is in the conformal window or
remains in the spontaneously chiral symmetry breaking phase, one needs
to know how the above picture of continuum limits would change if the
underlying continuum theory has a non-trivial IR fixed point.  In the
following discussion, we assume the lattice is infinitely large, $V
\rightarrow \infty$.  Since the weak coupling limit of the theory is
asymptotically free, the limit with $\beta \rightarrow \infty$ and
$m_q \rightarrow 0$ is a valid continuum limit representing the true
IR conformal continuum theory.  It is conceivable that any non-zero
quark mass term would break the conformal symmetry and the
renormalized theory would lose the IR fixed point.  If we keep
$m_q=0$, and vary $\beta$ away from the weak coupling limit, since the
renormalized coupling remains the fixed-point value, $g_\mathrm{IR}$,
the physics governed by the IR fixed point does not change, and
neither do the lattice beta functions.  Thus the theory remains IR
conformal with a vanishing lattice spacing, $a=0$.  In terms of
\eref{eq:RG}, any low-energy observable $P$ remains under control of
the IR fixed point, and is independent of the bare coupling, $g$, such
that increasing $g$ from $0$ does not need to be compensated by
increasing lattice spacing, $a$.  In this case, \eref{eq:RG} is
self-consistently satisfied with $a=0$ for $\hat{m}=0$ and $0 \le g
\le g_\mathrm{IR}$.  In conclusion, shown in \fref{bulk_trans}(b), in
the infinite volume limit, any point along the line with $m_q=0$ and
$\beta\ge6/g^2_\mathrm{IR}$ is a lattice critical point, and shares
the same continuum limit as the weak coupling limit, which describes
the same conformal theory at low energies.  On the other hand, if $m_q
\ne 0$, the theory is no longer conformal, and lattice artifacts are
present with $a\ne0$, which can be controlled by the bare coupling,
$\beta$.  This proposed scenario appears consistent with the data and
we are investigating it further.

\section{Mass ratios and universality}

\begin{figure}
  \centering
  \parbox{0.485\textwidth}{\includegraphics{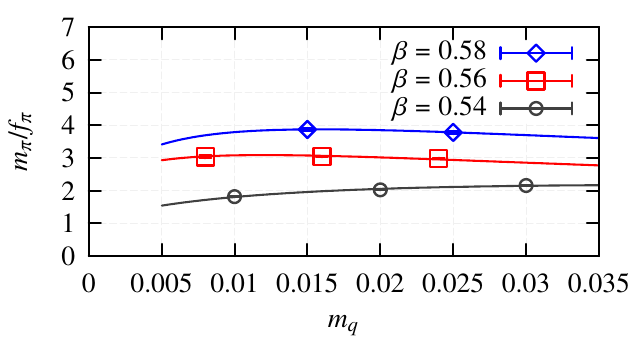}
    \figsubcap{(a) 8 flavors}}
  \hfill
  \parbox{0.485\textwidth}{\includegraphics{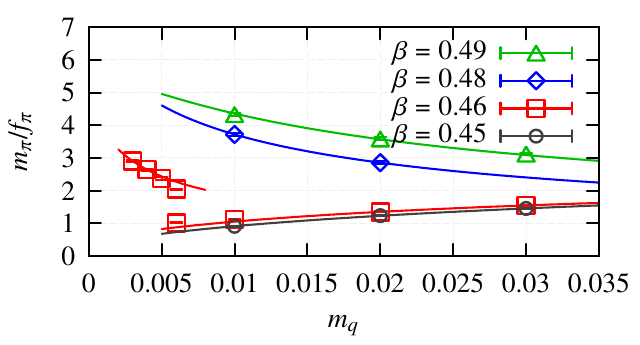}
    \figsubcap{(b) 12 flavors}}
  \caption{\label{mpi_fpi}Scaling of constant physics by
    $m_\pi/f_\pi$}
\end{figure}

The dimensionless quantity, $m_\pi/f_\pi$, is shown in \fref{mpi_fpi}
versus the bare input quark mass, for both 8 flavors (left panel) and
12 flavors (right panel), respectively.  We have drawn simple
polynomial fit curves on top of our data.  The discontinuity at
$m_q=0.006$ and $\beta=0.46$ with 12 flavors represents the bulk
transition.  No bulk transition has been observed within our
simulation parameters with 8 flavors.

We observed that, at strong couplings, $\beta=0.54$ and $\beta=0.45$
for 8 and 12 flavors respectively, the value of $m_\pi/f_\pi$
decreases as $m_q$ decreases.  This behavior is expected in a physical
QCD simulation, where chiral symmetry is spontaneously broken at
$m_q\rightarrow0$, and $m_\pi$ vanishes while $f_\pi$ remains finite
in the limit.  On the contrary, at weaker couplings, $\beta\ge0.56$
for 8 flavors, and $\beta\ge0.46$ after the bulk transition for 12
flavors, the $m_\pi/f_\pi$ increases as $m_q$ decreases.  With 8
flavors, $m_\pi/f_\pi$ is almost constant, and the fitted curves start
to bend towards zero at small $m_q$, while the fitted curves with 12
flavors continues to grow in the parameter range we have simulated.
Despite this difference, at weak couplings, $m_\pi/f_\pi$ with 12
flavors resembles that with 8 flavors at relatively large
$m_q\gtrsim0.015$.

If we look at the constant physics naively with fixed $m_\pi/f_\pi$ at
weak couplings of the 12-flavor simulations, in order to reach the
massless quark limit, we have to decrease $\beta$ and go to stronger
couplings.  This is in contrast with the continuum limit discussed in
the last section, and thus we cannot proceed with this procedure.  The
reason for this could be: a) lattice artifacts; b) physical behaviors.

\begin{figure}
  \centering
  \includegraphics{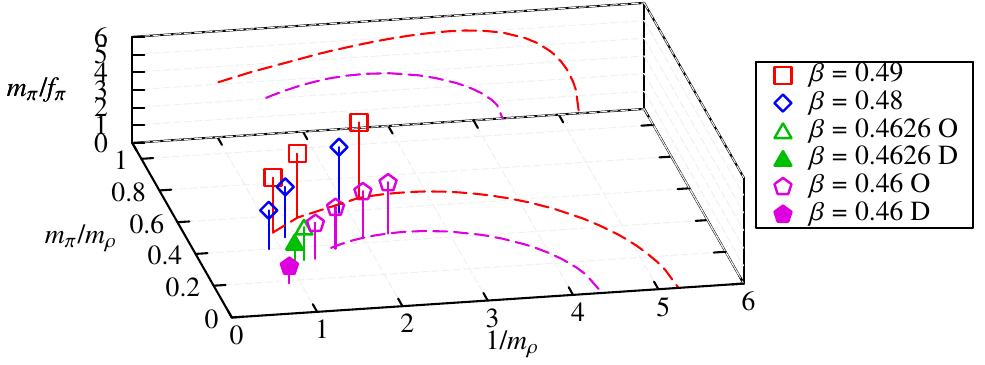}
  \caption{\label{phase_diagram_3d_jin}Diagram of dimensionless ratios
  versus typical lattice correlation length}
\end{figure}

We have assembled values of both $m_\pi/f_\pi$ and $m_\pi/m_\rho$ from
our 12-flavor simulations and plotted them against a typical lattice
correlation length $1/m_\rho$ in \fref{phase_diagram_3d_jin}.  In the
key of the figure, for ensembles with fixed $\beta$ across both sides
of the first order bulk transition, weak coupling side
(ordered-started) and strong coupling side (disordered-started) are
labeled with O and D, respectively.  The critical endpoint is located
near the data points with $\beta=0.4626$.  In the figure, the length
of the vertical line attached to each symbol represents $m_\pi/f_\pi$,
and its position on the bottom layer shows the value of $m_\pi/m_\rho$
versus the corresponding $1/m_\rho$.  Dashed lines are polynomial
extrapolations inspired by spontaneous-chiral-symmetry-breaking
(S$\chi$SB) behaviors,
\begin{align}
m_\pi^2 &= c_0 m_q (1 + c_1 m_q)\,, \\
m_\rho \text{ or } f_\pi &= b_0 + b_1 m_q + b_2m_q^2\,,
\end{align}
acted on each individual channel separately.  Assuming S$\chi$SB
behaviors, $1/m_\rho$ needs to be increasingly larger, with the
coupling weakening, to observe the ratios decrease as a result of
vanishing $m_\pi$.  On the other hand, if the theory does indeed have
an IR fixed point, ratios in the infinite volume limit should approach
constant values with different couplings, as $1/m_\rho \rightarrow
\infty$, while all the dimensionful quantities vanish in the conformal
limit.  Therefore, smaller quark masses would have stronger
differentiating power to tell the true behavior of the system.

\begin{figure}
  \centering
  \includegraphics{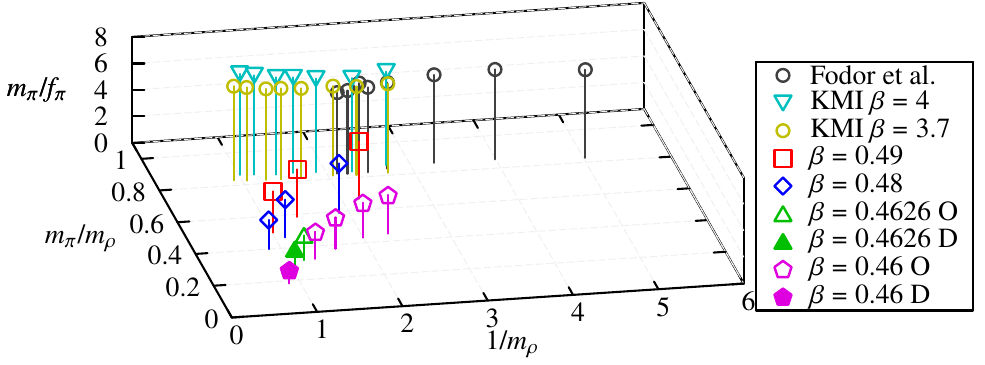}
  \caption{\label{phase_diagram_3d}Comparing dimensionless ratios with
    results from Fodor et al.\cite{Fodor:2011tu}\ and
    LatKMI\cite{Aoki:2012eq}}
\end{figure}

To compare our results with results from other groups, we have
included recent 12-flavor results\footnote{Results from the largest
  volume available are taken, and the finite volume effects are
  small.} from Fodor et al.\cite{Fodor:2011tu}\ and
LatKMI\cite{Aoki:2012eq},\footnote{Values of $f_\pi$ from LatKMI are
  normalized by dividing $\sqrt{2}$.} in \autoref{phase_diagram_3d}.
Although results of different groups are from different lattice
actions, there is the universal behavior to be learned from the
similar behaviors within the parameter space that has been explored.
Apparently, both ratios, $m_\pi/f_\pi$ and $m_\pi/m_\rho$ are not
changing noticeably as $1/m_\rho$ increases with fixed $\beta$ within
the parameter space limited by the computing resources.  However, this
behavior is not different from the behavior of $m_\pi/f_\pi$, shown in
\fref{mpi_fpi}(a), for a system with 8 flavors.  On the other hand,
the ratios do seem to approach a universal continuum limit at
$1/m_\rho \rightarrow \infty$, if the tendency continues without
alteration in the region with small quark masses.

\section{Summary}

We have located a lattice artifact, critical endpoint, where the
scalar correlation length diverges, for 12-flavor QCD with naive
staggered fermions and the DBW2 gauge action.  In the weak coupling
side of the lattice artifact bulk transition, we see confined, massive
particle states for all simulations done to date.  As the input quark mass
decreases at a fixed lattice coupling, the hadronic scales ($m_\rho$
and $f_\pi$) decrease much more rapidly than the pion mass, such that
$m_\pi/f_\pi$ or $m_\pi/m_\rho$ increases on the weak coupling side of
the bulk transition.  The lattice input quark mass at the critical
endpoint of the lattice artifact transition gives us a hint of the
size of quark masses, at that lattice coupling, where the vanishingly
small scalar mass can wildly distort continuum physics.

We have argued that an exact same continuum limit of an asymptotically
free, IR conformal theory can be approached at the massless quark
limit and the infinite volume limit, with any lattice bare coupling
that is weaker than the renormalized coupling at the IR fixed point.

It may not be surprising that much lighter quark masses are required
in lattice simulations for the S$\chi$SB behavior to be numerically
indisputable.  Nevertheless, how to effectively scale towards the
continuum limit within our current computational resources remains a
question for both S$\chi$SB and IR conformal scenarios.

\section*{Acknowledgments}
X.-Y.~Jin would like to thank the organizers of SCGT12 for the
invitation to this wonderful workshop.  Our calculations were done on
the QCDOC and NY Blue at BNL\@.  X.-Y.~Jin wants to express the
sincere gratitude especially towards all the brilliant minds behind
the QCDOC project.  This research utilized resources at the New York
Center for Computational Sciences at Stony Brook University/Brookhaven
National Laboratory which is supported by the U.S. Department of
Energy under Contract No.~DE-FG02-92ER40699 and by the State of New
York.

\bibliographystyle{ws-procs975x65}
\bibliography{ref}

\end{document}